\documentstyle[aps,twocolumn,epsfig]{revtex}
\begin{document}
\newcommand{\ve}[1]{\mbox{\boldmath $#1$}}
\twocolumn[\hsize\textwidth\columnwidth\hsize
\csname@twocolumnfalse%
\endcsname

\draft

\title{Bose-Einstein condensation of excitons in Cu$_2$O}
\author{G. M. Kavoulakis}
\date{\today}
\address{Royal Institute of Technology, Lindstedtsv\"agen 24,
         S-10044 Stockholm, Sweden}

\maketitle

\begin{abstract}

We present a parameter-free model that estimates the density of excitons
in Cu$_2$O, related to experiments that have tried to create an excitonic
Bose-Einstein condensate. Our study demonstrates that the triplet-state
excitons move along adiabats and obey classical statistics, while the
singlet-state excitons are a possible candidate for forming a
Bose-Einstein condensate. Finally we show that the results of this
study do not change qualitatively in a two-dimensional exciton gas,
which can be realized in a quantum well.

\end{abstract}
 
\vskip0.5pc]

\section{Introduction}

The field of Bose-Einstein condensation \cite{GSS} has attracted much 
attention in recent years, specially in the context of trapped alkali-metal
atoms \cite{RMP}. Much experimental effort has also been put in creating a
Bose-Einstein condensate \cite{MS} of excitons \cite{Knox}, both in 
three-dimensional crystals \cite{David,Andre,SW,Jim,Keith}, and in 
quantum wells \cite{QW}.

Excitons are bound states of electrons and holes, which form in semiconductors
after the electrons get excited from the valence to the conduction band
with use of some external probe, a laser beam of light, for example.
As long as the mean exciton-exciton spacing, which is of order 
$n^{-1/3}$, with $n$ being the exciton density, is much larger 
than the exciton Bohr radius $a_B$, one can treat the excitons
as point-like particles that behave as bosons. 
On the other hand for the exciton gas to form a Bose-Einstein 
condensate, the density has to be sufficiently high
and the exciton temperature $T$ has to be sufficiently low, in order 
for the thermal de Broglie wavelength of the excitons to be comparable to the
mean exciton-exciton spacing. This condition gives rise to
the formula connecting the critical density $n_c$ for Bose-Einstein
condensation to the critical temperature $T_c$, which in three dimensions
and for a homogeneous ideal gas gives, 
\begin{eqnarray}
   n_c &=& g \zeta(3/2)
  \left( \frac {m k_B} {2 \pi \hbar^2} \right)^{3/2} T_c^{3/2}, 
\label{cond}
\end{eqnarray}
where $g$ is the (spin) degeneracy, and $m$ is the exciton mass.

Cu$_2$O is a material often used to perform such experiments
with excitons \cite{David,Andre,SW,Jim,Keith}. 
The goal is, for typical temperatures
determined by the lattice temperature (i.e., liquid helium 
temperatures), or even somewhat higher, to achieve densities 
that are in the range we described above: high enough for
the quantum effects to show up, but low enough so that the excitons
still behave as point-like particles, and therefore as bosons.
In terms of real numbers, assuming, for example, that the 
temperature is $T = 2$ K, the corresponding density has to be 
of order $10^{17}$ cm$^{-3}$. Under such conditions the mean
exciton-exciton spacing is a few hundreds angstroms, which is
much larger than the exciton Bohr radius ($a_B \approx 5.29$ \AA\, 
in Cu$_2$O \cite{KCB}).

In these experiments spectroscopic methods have  
implied that the spin-triplet orthoexcitons 
are highly degenerate \cite{David,Andre,SW} moving parallel to the phase 
boundary. More specifically their chemical potential which is extracted 
by fitting the phonon-assisted recombination lines to Bose-Einstein
distributions is found to be very close to zero. In addition the 
same method implies that the spin-singlet paraexcitons have formed a
Bose-Einstein condensate \cite{Jim}. The lines parallel to the phase
boundary given by Eq.\,(\ref{cond}) are adiabats, i.e., along them the entropy
per exciton is constant, which is crucial in the discussion that
follows below. However, more recent studies \cite{Keith} of the same system
have estimated the number of excitons created and the volume they
occupy, and have shown a clear discrepancy, since the densities extracted
are two to three orders of magnitude lower; for such low densities, even at the
lowest temperatures, the exciton statistics should be completely classical.

In the present study we give theoretical estimates for the density
of the ortho and paraexcitons as function of their temperature, based on a 
parameter-free model. We also give suggestions
and predictions for future experimental studies, which would put an end 
to the discrepancies described above. We also examine the conditions 
under which a Bose-Einstein condensate would form, since, as we 
show, this is possible for the paraexcitons. Finally, in connection 
with excitons in quantum wells, we demonstrate that our conclusions 
are unaffected for a two-dimensional exciton gas.

This paper is organized in the following way: we give 
in Sec.\,II a simple and analytical argument for the fact that
the orthoexcitons move along adiabats \cite{David,Andre,SW}. 
We estimate the density of the orthoexcitons and from that we extract
the paraexciton density in Sec.\,III, discussing the long-time
behavior of the paraexcitons. In Sec.\,IV we analyze the absorption
spectrum of infrared electromagnetic radiation inducing transitions from the
$1s$ to the $2p$ level \cite{JK} under realistic conditions,
and in Sec.\,V we investigate the behavior of a quasi-two-dimensional exciton
gas. Finally we examine the two dominant mechanisms that determine the
behavior of the orthoexcitons in Sec.\,VI, and summarize our 
results in Sec.\,VII.

\section{Adiabatic behavior of the orthoexcitons}

Let us first of all assume that the orthoexcitons and the paraexcitons 
occupy the same volume inside the crystal, and that they are described 
by a common temperature.
Our claim is that the orthoexcitons in this system move along
adiabats as a result of the balance between two competing mechanisms:
a heating mechanism that comes from an ortho-plus-ortho to 
para-plus-para conversion mechanism \cite{KM}, and an acoustic-phonon cooling 
process \cite{KBW}. A similar argument was used in Ref.\,\cite{KBW}, 
but the heating process was assumed to be an Auger mechanism \cite{Snoke,KB}.
However, in view of the more recent experimental
measurements \cite{Keith} and theoretical calculations \cite{KM},
this Auger mechanism has a negligible rate. The model we present here
is essentially {\it free of adjustable parameters}, which makes it very 
reliable \cite{notee}.

Let us therefore give the general argument, and then examine each mechanism 
separately. We know from elementary thermodynamics that
\begin{eqnarray}
 d U_o = T d S_o - p_o d V + \mu_o d N_o,
\label{therm}
\end{eqnarray}
where $U_i$ is the internal energy, $S_i$ is the entropy, $p_i$ is the  
pressure, $\mu_i$ is chemical potential, $N_i$ is the number 
of species $i$, and $V$ is the volume of the exciton gas. 

Solving Eq.\,(\ref{therm}) in terms of $T d S_o$, and expanding the 
differential $d(S_o/N_o)$ we get
\begin{eqnarray}
  T d \left(\frac {S_o}{N_o} \right) = \frac 1 N_o 
 \left[ d U_o - \left(\mu_o + \frac {T S_o} {N_o} \right) d N_o 
+ p_o dV \right].
\label{thermm}
\end{eqnarray}
We show below that the orthoexcitons are in the classical regime,
and for an ideal monatomic classical gas $\mu_o + T S_o/N_o = 5 k_B T/2$,
and also $p_o V = N_o k_B T$. Therefore, Eq.\,(\ref{thermm}) can be 
written as
\begin{eqnarray}
  T d \left(\frac {S_o}{N_o} \right) = \frac 1 {N_o}
 \left( d U_o - \frac {5} {2} k_B T \,\, d N_o + n_o k_B T dV \right).
\label{thermmf}
\end{eqnarray}
The differential $d U_o$ is determined by the two mechanisms we mentioned.
More specifically, as discussed in Sec.\,VI, the heating rate is equal to 
$a n_o$, $a$ being a constant, i.e., it scales linearly with 
the orthoexciton density $n_o$, whereas the acoustic-phonon cooling 
rate is equal to $-b T^{3/2}$ (for $T \gg T_l$, where $T_l$ is the 
lattice temperature), $b$ being a constant, i.e., it
scales as $T^{3/2}$. Therefore Eq.\,(\ref{thermmf}) implies that 
\begin{equation}
   T \frac d {dt} \left(\frac {S_o}{N_o} \right) \approx 
   a n_o - b T^{3/2} - \frac {5} {2}  
 \frac {k_B T} {N_o} \frac {d N_o} {dt} + \frac {k_B T} {V} \frac {dV} {dt}.
\label{thermmfin}
\end{equation}
Let us now examine the terms on the right side of Eq.\,(\ref{thermmfin}).
It turns out that the first two are the dominant ones. 
\noindent
\begin{figure}
\begin{center}
\epsfig{file=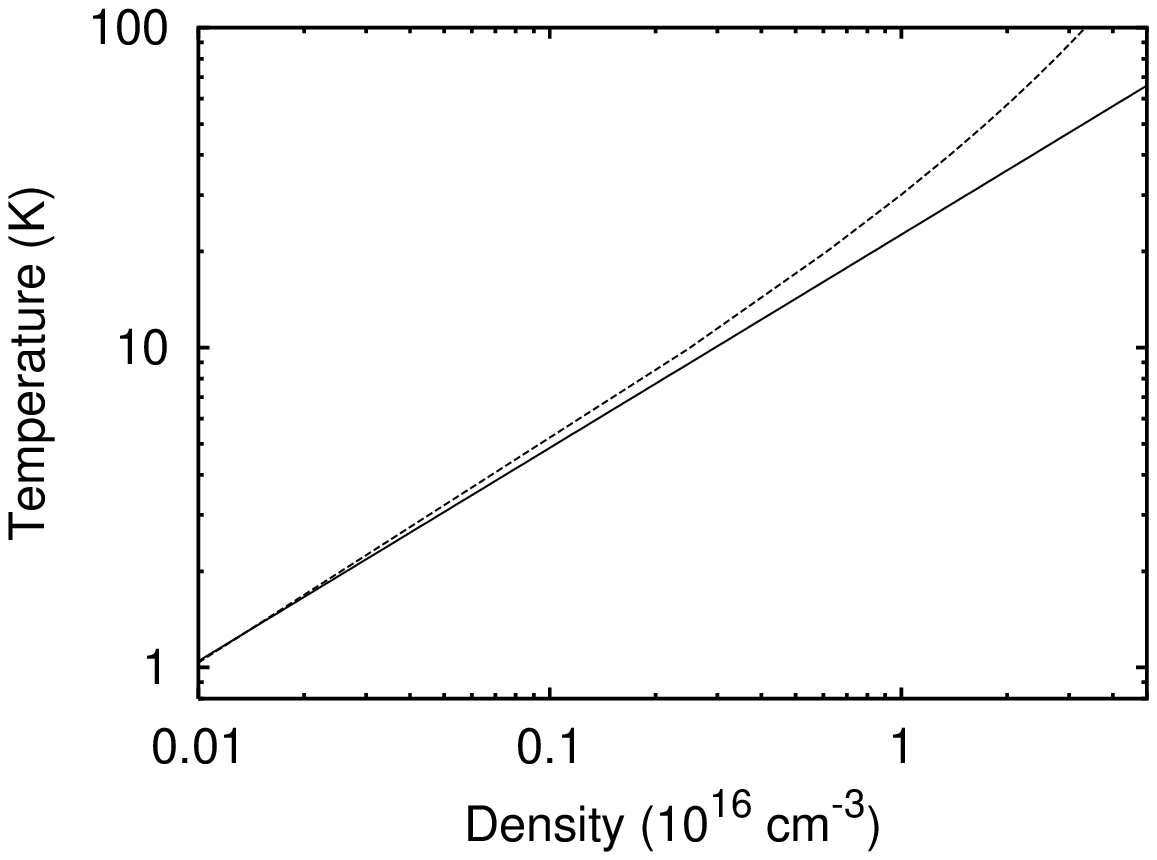,width=8cm,height=8cm,angle=0}
\vskip1pc
\begin{caption}
{The orthoexciton trajectories on the density-temperature plane. The
solid line gives the root of the polynomial on the right side of
Eq.\,(\ref{new}), whereas the dashed line gives the root of the
polynomial on the right side of Eq.\,(\ref{thermmfinn}).}
\end{caption}
\end{center}
\label{FIG1}
\end{figure}
\noindent
More specifically
as we show in Sec.\,VI the order of magnitude of these two terms
(i.e., the typical energy-exchange rates that come from the phonon
cooling and the ortho-to-para exchange heating) is $\agt 10$ meV/ns.
On the other hand, concerning the last term that comes from the expansion
of the orthoexcitons, a detailed study of this problem has given
that the diffusion constant is $D \sim 10^3$ cm$^2$/s \cite{Tr}. Even at
the initial stage of the expansion of the cloud when its radius is 
$R \agt 10$ $\mu$m, $dV/(V dt) \sim D/R^2 \alt 1$ ns$^{-1}$; thus this
term is $\alt 1$ meV/ns for $T \approx 10$ K. Concerning the third term on the
right side of Eq.\,(\ref{thermmfin}) one has to distinguish between
two different cases corresponding to two different experimental conditions:
For long-pulse laser excitation, of order of a few tens of nanoseconds, the 
population of orthoexcitons is essentially determined by the laser
and it follows its profile (see, for example Fig.\,4(a) of Ref.\,\cite{Jim}).
Therefore the characteristic timescale $\tau_L$ that is related to this 
mechanism is a few tens of nanoseconds, and the corresponding term
entering Eq.\,(\ref{thermmfin}) is $\sim k_B T/\tau_L$, or $\alt$ 0.1 meV/ns,
and thus negligible. Under these conditions Eq.\,(\ref{thermmfin}) takes the 
form
\begin{eqnarray}
    T \frac d {dt} \left(\frac {S_o}{N_o} \right) \approx
    a n_o - b T^{3/2}.
\label{new}
\end{eqnarray}
Equation (\ref{new}) is one of the basic results of this study. 
To see what this equation implies, it is instructive to give the 
two following formulas connecting the entropy, number, density, and
temperature of the orthoexcitons (which are valid for any ideal Bose gas) 
\cite{Pathria}
\begin{eqnarray}
  \frac {S_o} {N_o k_B} = \frac 5 2 \frac {g_{5/2}(z_o)} {g_{3/2}(z_o)} 
 - \ln z_o;
\label{so}
\end{eqnarray}
\begin{eqnarray}
  \frac {n_o} {T^{3/2}} = 
 \left( \frac {m k_B} {2 \pi \hbar^2} \right)^{3/2} g_{3/2}(z_o),
\label{soo}
\end{eqnarray}
where $z_o=e^{\mu_o/k_B T}$ and
\begin{eqnarray}
g_n(z) \equiv \frac 1 {\Gamma(n)} \int_o^{\infty}
\frac {x^{n-1} dx} {z^{-1} e^x - 1},
\label{C2}
\end{eqnarray}
with $\Gamma(n)$ being the gamma function. From the above equations
one sees that along lines of constant $\mu_o/k_B T$, i.e., of constant $z_o$, 
$S_o/N_o$ and $n_o/T^{3/2}$ are both constant. Therefore along adiabats, i.e., 
along lines with constant $S_o/N_o$, $n_o \propto T^{3/2}$ and according to
Eq.\,(\ref{new}), the orthoexcitons approach an adiabat,
and more specifically the one for which $n_o = (a/b) T^{3/2}$.
This is a line of ``equilibrium" for the orthoexcitons, which comes
as a balance between the heating and the cooling mechanisms.
Using the values for the constants $a$ and $b$ given in Sec.\,VI, this
adiabat is [see the solid line in Fig.\,1]
\begin{eqnarray}
  n_o(10^{16} {\rm cm}^{-3}) \approx 10^{-2} T^{3/2}(\rm {K}),
\label{adline}
\end{eqnarray}
where the notation $n_o(10^{16} \rm{cm}^{-3})$ means that the density
is to be measured in units of $10^{16} \rm{cm}^{-3}$, and correspondingly
the temperature in Kelvin. The coefficient $10^{-2}$ in Eq.\,(\ref{adline})
should be compared with $\approx 3.3$ given by Eq.\,(\ref{cond}), which
gives the phase boundary, and therefore this adiabat is in the regime
where the excitons should not exhibit any sign of quantum degeneracy, since
$-\mu_o/k_B T \gg 1$. For a typical exciton temperature of order 20 K, the above
equation implies that the orthoexciton density is of order $10^{16}$ cm$^{-3}$,
which is consistent with that estimated experimentally in 
Refs.\,\cite{Keith}. 

One can also study the dynamics of orthoexcitons. To do this,
it is convenient to introduce the dimensionless quantity 
$\alpha=-\mu_o/k_B T$ and rewrite Eq.\,(\ref{new}) in the form
\cite{KBW} 
\begin{eqnarray}
  \frac {d \alpha} {dt} = - \frac {\alpha - \alpha_*} {\tau_*},
\label{newexpr}
\end{eqnarray}
where $\alpha_*$ is the value of $\alpha$ along the specific adiabat
with $n_o = (a/b) T^{3/2}$. Also
\begin{eqnarray}
  \tau_* = - \frac T b \left[
 \frac {\partial (S_o/N_o)} {\partial n_o} \right]_T \approx
 \frac {5 \zeta(5/2)} {2 \zeta(3/2)} \frac {k_B} {a T^{1/2}}.
\label{newexprtau}
\end{eqnarray}
Equation (\ref{newexpr}) has to be solved along with the rate equations
describing the change in the exciton density of each species,
\begin{eqnarray}
    \frac {d n_o} {d t} = G_o(t) - \frac {n_o} {\tau_{o,p}} 
  + \frac {n_p} {\tau_{p,o}} - \frac {n_o} {\tau_{lo}}, 
\label{21} \\
 \frac {d n_p} {d t} = G_p(t) - \frac {n_p} {\tau_{p,o}}
+ \frac {n_o} {\tau_{o,p}} - \frac {n_p} {\tau_{lp}},
\label{22}
\end{eqnarray}
where $G_i(t)$ is the laser production rate of excitons, $\tau_{li}$ are 
the intrinsic radiative lifetimes, $\tau_{o,p}^{-1}$ is the rate
for the ortho-to-para conversion process, and $\tau_{p,o}^{-1}$
is the rate for the (reverse) para-to-ortho conversion process. 
As shown in Ref.\,\cite{KM}, $\tau_{o,p}^{-1} \approx c n_o$ 
($c$ is a constant that is given in Sec.\,VI), whereas as explained
in Sec.\,VI the reverse process is thermally supressed, $\tau_{p,o}^{-1}
\ll \tau_{o,p}^{-1}$. In addition the radiative lifetimes of excitons
are very long (longer than microseconds), and thus Eqs.\,(\ref{22}) and 
(\ref{23}) take the simple form
\begin{eqnarray}
    \frac {d n_o} {d t} \approx G_o(t) - \frac {n_o} {\tau_{o,p}},
\label{23} \\
 \frac {d n_p} {d t} \approx G_p(t) + \frac {n_o} {\tau_{o,p}}.
\label{24}
\end{eqnarray}
Equations (\ref{23}) and (\ref{24}), viewed as densities being functions of  
time, do not depend on the exciton temperature and can be integrated.
Therefore Eqs.\,(\ref{newexpr}), (\ref{23}), and (\ref{24}) can be integrated  
to give $n_o(t)$, $n_p(t)$, and $T(t)$ \cite{KBW}. However we are interested 
in $n_o(T)$, i.e., in the dependence of $n_o$ on the exciton temperature 
$T$, and that is given directly by Eq.\,(\ref{newexpr}), which implies
that the orthoexcitons approach the adiabat with $\alpha=\alpha_*$ on a
timescale of order $\tau_*$, which is of order of a few tens of picoseconds
under typical exciton temperatures. 

In another class of experiments short laser pulses (on the order of 
100 ps) have been used in order to excite the crystal. In this case 
one must include the third term on the right side of Eq.\,(\ref{thermmfin}) 
which, however, does not change appreciably our basic results. More
precisely Eq.\,(\ref{thermmfin}) takes the following form in this case
\begin{eqnarray}
    T \frac d {dt} \left(\frac {S_o}{N_o} \right) \approx
    a n_o - b T^{3/2} - \frac {5 c} {2} k_B T n_o.
\label{thermmfinn}
\end{eqnarray}
The dashed line in Fig.\,1 shows the root of the polynomial  
$a n_o - b T^{3/2} - (5 c/2) k_B T n_o = 0$. Although the deviation from
the adiabatic behavior is not substantial, it is in agreement with the  
experimental observations, as one can see in Fig.\,2 of Ref.\,\cite{SW}, for 
example. The open circles in this graph correspond to short laser pulse
excitation ($\sim 100$ ps) and they show some small deviation from the adiabats,
as opposed to the solid circles (corresponding to long pulses, $\sim 10$ ns) 
that follow very closely an adiabat.

From the discussion we have presented up to now there is
a contradiction: on the one hand spectroscopically
the orthoexcitons show a high degree of quantum degeneracy 
\cite{David,Andre,SW,Jim}. On the other hand, 
we argued that the orthoexcitons should not show any sign of quantum 
degeneracy (under equilibrium conditions), in agreement with the 
experimental data presented in Refs.\,\cite{Keith}. We will not address
this issue here, which is an open question for future studies.

\section{Quantum degeneracy of paraexcitons}

Let us now turn to the paraexcitons. Their radiative lifetime has been
determined to be on the order of milliseconds \cite{Keith,Karl}, which allows
them to form a cold gas, with a temperature very close to $T_l$ on these
timescales. Since on such timescales and for such low temperatures, all the
excitons will have converted to paraexcitons, they are also expected to 
establish a relatively high density, which we estimate now, given
the orthoexciton density we got earlier. Statistically due to the
multiplicity of the orthoexcitons, one expects that $N_o/N_p=3$. Therefore
the paraexciton density (assuming that all the orthoexcitons get
converted into paraexcitons) is four times the one we found earlier.
However, as we argued the temperature is expected to get very close to that
of the lattice for late times, i.e., for times much larger than the 
exciton - acoustic phonon scattering time (on these timescales the 
ortho-to-para conversion process will have converted all the orthoexcitons
to paraexcitons and there will be no heating due to this mechanism).  
Therefore for an initial exciton temperature $T_i=40$ K, the paraexciton 
density can get up to 10$^{17} T_i^{3/2}(40 \,{\rm K})$ cm$^{-3}$, as
Eq.\,(\ref{adline}) implies, where $T_i$ is to be measured in units
of 40 K. Even for a density of $10^{17}$ cm$^{-3}$
the critical temperature for Bose-Einstein condensation of the paraexcitons
is approximately 2 K, i.e., liquid-helium temperatures. One important 
conclusion of this analysis is, therefore, that the paraexcitons in the
experiments that have been performed \cite{David,Andre,SW,Jim} should 
be very close to the phase boundary, or they might have even crossed it already.
In the experiment of Ref.\,\cite{Jim} the paraexcitons were reported 
to have crossed the phase boundary for Bose-Einstein condensation.
The whole analysis was based on spectroscopically analyzing the (only) 
phonon-assisted recombination line of paraexcitons, which is very weak,
and is close to other, much stronger lines. An alternative way of 
probing experimentally the degree of quantum degeneracy of paraexcitons 
is presented in the following section.

\section{Absorption spectrum of radiation inducing internal transitions: 
Probing the quantum degeneracy}

In a possible experiment \cite{Karl} that has been proposed and
has been studied theoretically in Ref.\,\cite{JK}, it was
shown that the absorption spectrum of infrared radiation
inducing internal transitions of the excitons from the $1s$ to
the $2p$ level is very sensitive to the degree of quantum degeneracy 
of the gas. Therefore, in such an experiment one should be able to observe the
contribution of the orthoexcitons and the paraexcitons to the absorption 
separately, with an energy separation of order $\Delta E$ (assuming
that the width of each distribution is of order $k_B T \ll \Delta E$).
Here $\Delta E$ is the energy splitting between the orthoexcitons (lying 
higher than) the paraexcitons due to the exchange interaction \cite{KCB}.
At the zone center $\Delta E \approx 12$ meV in Cu$_2$O,  which corresponds to 
approximately 150 K. In a temporal study of this experiment, following
a short laser-pulse excitation (like the one already used of a few hundred
picoseconds) the contribution of the orthoexcitons to the absorption 
of infrared radiation would vanish within a timescale of order $\tau_{o,p}$,
i.e., of order nanoseconds under typical conditions. On the other hand
the contribution of the paraexcitons to the absorption would last much longer,
on a timescale of order of their radiative lifetime, i.e., 
milliseconds \cite{Keith,Karl}. As shown in Ref.\,\cite{JK}
the appearance of two distinct peaks in the absorption spectrum 
of paraexcitons would signal the presence of a Bose-Einstein condensate, since
in the condensed phase one deals with a two-component system
and the two peaks would indicate the two different collective
modes of it. But even if the paraexcitons have not crossed the
phase boundary, but they are highly degenerate, that would still
show up clearly in the absorption spectrum \cite{JK}. An additional
advantage of this method is that it does not depend on the
strength of the radiative recombination lines (which is very weak for
the paraexcitons, as we mentioned in the preceding section).

It is interesting that if the lattice temperature increases and
becomes of order the orthoexciton-paraexciton splitting, that would result 
in more or less equal rates for the ortho-to-para and the para-to-ortho
conversion processes, and would decrease the paraexciton density
substantially \cite{Mick}. In such a case the orthoexciton lifetime 
would be determined by the radiative lifetime, i.e., microseconds 
\cite{Keith,Karl}. Therefore keeping $T_l$ as low as 
possible is very crucial, since it (i) enhances the paraexciton density,
and (ii) determines the temperature of the paraexcitons for late times.

\section{A two-dimensional exciton gas}

As a final remark we comment on the possibility of confining
the excitons and creating a quasi-two-dimensional gas, like in
quantum wells \cite{QW}. As we show in Sec.\,VI in two dimensions the 
acoustic-phonon cooling rate scales linearly with the temperature
for $T \gg T_l$. On the other hand the ortho-to-para conversion
mechanism is unaffected by the dimensionality of the system.
Thus the argument we gave earlier in Eq.\,(\ref{new})
now takes the form
\begin{eqnarray}
    T \frac d {dt} \left(\frac {S_o}{N_o} \right) \approx
    a \sigma_o - b T,
\label{new2d}
\end{eqnarray}
where $\sigma_o$ is the orthoexciton surface density. For $S_o/N_o$
to be constant in two dimensions $\sigma_o \propto T$, and therefore
according to Eq.\,(\ref{new2d}) the orthoexcitons in Cu$_2$O would 
still move along an adiabat $\sigma_o = (b/a)T$. In addition a 
two-dimensional gas is expected to undergo a 
Kosterlitz-Thouless transition \cite{KT} to a superfluid state along
lines on which $\sigma_o \propto T$ and thus the orthoexcitons
would also move along the phase boundary for 
the Kosterlitz-Thouless transition without crossing it,
as opposed to the paraexcitons that would have the chance of
undergoing this phase transition.

\section{Ortho to para conversion and acoustic-phonon cooling}

Let us now examine in detail the two processes we mentioned earlier,
starting with the spin-exchange mechanism, where two orthoexcitons
collide, exchange their electrons or their holes,
resulting into two paraexcitons in the final states.
This process has been studied in detail in Ref.\,\cite{KM} and
the decay time $\tau_{o,p}$ has been calculated to be in very good 
agreement with experiment \cite{Keith}, with essentially no adjustable
parameter. The decay rate was shown to be given by
\begin{eqnarray}
   \tau_{o,p}^{-1} \approx 5 n_o(10^{16} \rm{cm}^{-3}) \, \rm{ns}^{-1}.
\label{numrr}
\end{eqnarray}
The reverse process is thermally supressed for temperatures much lower than
$\Delta E/k_B$. The ortho-to-para conversion is also a heating mechanism,
since there is a gain of energy $\Delta E$ per particle.
Therefore the heating rate due to this process is given by
\begin{eqnarray}
   \frac 1 {N_o} \left( \frac {\partial U_o} {\partial t} \right)_{o,p}= 
  \frac {\Delta E} {\tau_{o,p}} \approx 60 \, n_o(10^{16} \rm{cm}^{-3}) 
 \, \rm{meV}/\rm{ns}.
\label{heatr}
\end{eqnarray}

We conclude with the cooling process of excitons due to their
scattering with phonons. The excitons interact with the 
lattice vibrations, and since the lattice temperature $T_l$ is kept
low (i.e., a few K), the excitons, which in general
have a higher temperature, are cooled down. At the low energies we
consider here the process is dominated by collisions with acoustic 
phonons. The corresponding cooling rate per particle is given by 
\cite{KBW}
\begin{eqnarray}
   \frac 1 {N_o} \left( \frac {\partial U_o} {\partial t} \right)_{ph} \approx
  - 0.56 \, T^{3/2}({\rm K}) \left( 1 - \frac {T_l} T \right)
 \, \rm{meV}/\rm{ns}.
\label{coolr}
\end{eqnarray}
The above cooling rate scales as $T^{3/2}$ for $T \gg T_l$, with
the matrix element due to deformation potential theory, the 
acoustic-phonon energy, and the density of states in three 
dimensions each contributing a factor of $T^{1/2}$ (they scale linearly
with the momentum exchange). For a two-dimensional gas the density of
states is constant, and the cooling rate is proportional to $T$ is this case.

\section{SUMMARY}

In conclusion, based on a parameter-free model, we have given 
estimates for the density of the excitons in Cu$_2$O as a function
of temperature. The triplet state excitons are expected to 
move along adiabats, and to obey classical statistics. We have argued
that this adiabatic behavior is a result of the competition between
a spin-exchange heating mechanism, and an acoustic-phonon cooling
process. An open question that needs to be investigated is the chemical
potential of orthoexcitons, which, although they obey classical statistics,
is very close to zero. 

In addition, according to our model, the singlet-state excitons, are
a possible candidate for forming a Bose-Einstein condensate. Finally, 
we have shown that in a two-dimensional exciton gas the basic conclusions
of our study remain the same.

\vskip2pc

\centerline{\bf ACKNOWLEDGMENTS}

\vskip1.0pc

  I am grateful to G. Baym, M. G\"oppert, A. Jolk, K. Johnsen, M. J\"orger,
C. Klingshirn, and A. Mysyrowicz for useful discussions.


\begin{references}
\bibitem{notee} It has been argued that a phonon-assisted mechanism
   also converts orthoexcitons to paraexcitons \cite{KMA,STW}. We ignore
   this mechanism here, but since its rate scales as $T^{3/2}$, even
   if it is non-negligible, it does not change the results of the 
   present study, apart from pushing the orthoexcitons to an adiabat which
   is even further away from the phase boundary, and enhancing the paraexciton 
   density.
\end{references}
\end{document}